\documentclass[twocolumn, showpacs, amsmath, amssymb, superscriptaddress]{revtex4-1}

\usepackage{graphicx}
\usepackage{txfonts}

\begin{document}

\title{Field-induced Confined States in Graphene}

\author{Satoshi Moriyama}
\email{Corresponding author. \\ Email address: MORIYAMA.Satoshi@nims.go.jp}
\affiliation{
International Center for Materials Nanoarchitectonics (WPI-MANA), National Institute for Materials Science (NIMS), 1-1 Namiki, Tsukuba, Ibaraki 305-0044, Japan}%
\author{Yoshifumi Morita}
\affiliation{
Faculty of Engineering, Gunma University, Kiryu, Gunma 376-8515, Japan}%
\author{Eiichiro Watanabe}
\affiliation{
Nanotechnology Innovation Station, NIMS, 1-2-1 Sengen, Tsukuba, Ibaraki 305-0047, Japan
}%
\author{Daiju Tsuya}
\affiliation{
Nanotechnology Innovation Station, NIMS, 1-2-1 Sengen, Tsukuba, Ibaraki 305-0047, Japan
}%

\date{\today}

\begin{abstract}
We report an approach to confine the carriers in single-layer graphene, which leads to quantum devices with field-induced quantum confinement. We demonstrated that the Coulomb-blockade effect evolves under a uniform magnetic field perpendicular to the graphene device. Our experimental results show that field-induced quantum dots are realized in graphene, and a quantum confinement-deconfinement transition is switched by the magnetic field.
\end{abstract}

\maketitle


Graphene consists of a single layer of carbon atoms and has provided an attractive stage for studying novel two-dimensional electron gases (2DEGs) that exhibit a Dirac-particle-type gapless (`massless') linear (`relativistic') energy dispersion \cite{NovoselovScience04, NovoselovNature05,ZhangNature05}. Triggered by recent progress in graphene physics and technology, immense effort has been devoted to fabricate graphene nanodevices such as quantum wires or quantum dots (QDs). However, confining massless Dirac fermions in graphene is difficult due to Klein tunneling and the zero-band-gap electronic structure \cite{KatsnelsonNPhys06}. Therefore, although attempts have been made to design graphene devices using quantum confinement, they often suffer from severe design limitations. They basically consist of small quantum-dot islands, which confine electrons geometrically, to which narrow graphene-constrictions are connected \cite{StampferAPL08,PonomarenkoScience08,SchnezAPL09,LiuPRB09,GuttingerPRL09,MoriyamaNanoLett09,MolitorAPL09}. In this case, the device performance has been limited due to detailed constriction and edge orientation. Recently, several bottom-up processes to fabricate nanoribbons have been developed that control the edge structure at the atomic level \cite{LiScience08,CamposNanoLett09,WangNNanotech11}. 
But a lot of research is still needed to develop fabrication processes capable of controlling the atomic-scale edge orientation for a given geometry. 

To address this situation, we propose herein a new device structure, in which graphene mesoscopic islands are perfectly isolated and metallic contacts are directly deposited onto them without constrictions. Such a configuration is free from disturbances due to the structural fluctuation of constrictions and allows direct contact to the mesoscopic 2DEG system. We present an experimental demonstration of  a `field-induced' Coulomb-blockade effect and quantum confinement in the graphene device. This confinement is induced by both a uniform magnetic field perpendicular to the graphene sheet and an electrostatic surface-potential formed by the metal/graphene junction. 

The graphene samples were prepared by micromechanical cleavage of Kish graphite deposited on the surface of a silicon substrate with a 90 nm oxidized silicon. From these samples, we selected a number of graphene flakes by optical microscope contrast and Raman spectroscopy measurements \cite{FerrariPRL06, MoriyamaSTAM10}. Using these techniques, we confirmed that the samples used for the present study consisted of a single-layer graphene. The isolated-graphene-mesoscopic structures were patterned by electron-beam (EB) lithography using polymethyl-methacrylate (PMMA) resist as an etch mask, with $\text{O}_2$ reactive ion etching being used to etch away the unprotected graphene. Next, EB lithography was again used to deposit metal electrodes on the graphene mesoscopic structures to fabricate the source-drain contacts. Finally, the samples were annealed in Ar and $\text{H}_{2}$ (3 \%) at 300 ${}^\circ\text{C}$ for 5 min. 
Moreover, in order to use as a reference of our graphene flake, we also made a Hall-bar device on it. 
From the four-terminal measurement, we observed the integer quantum Hall effect of a single-layer graphene at low temperatures, 
and estimated the flake mobility to be $\approx$ 2,500 $\text{cm}^{2} / \text{Vs}$, the mean free path $\approx$ 30 nm 
at $10^{12} \text{cm}^{-2}$ hole- and electron-carrier densities \cite{MoriyamaMEX13}. 

\begin{figure}
\includegraphics[width=8cm, bb=0 0 786 821]{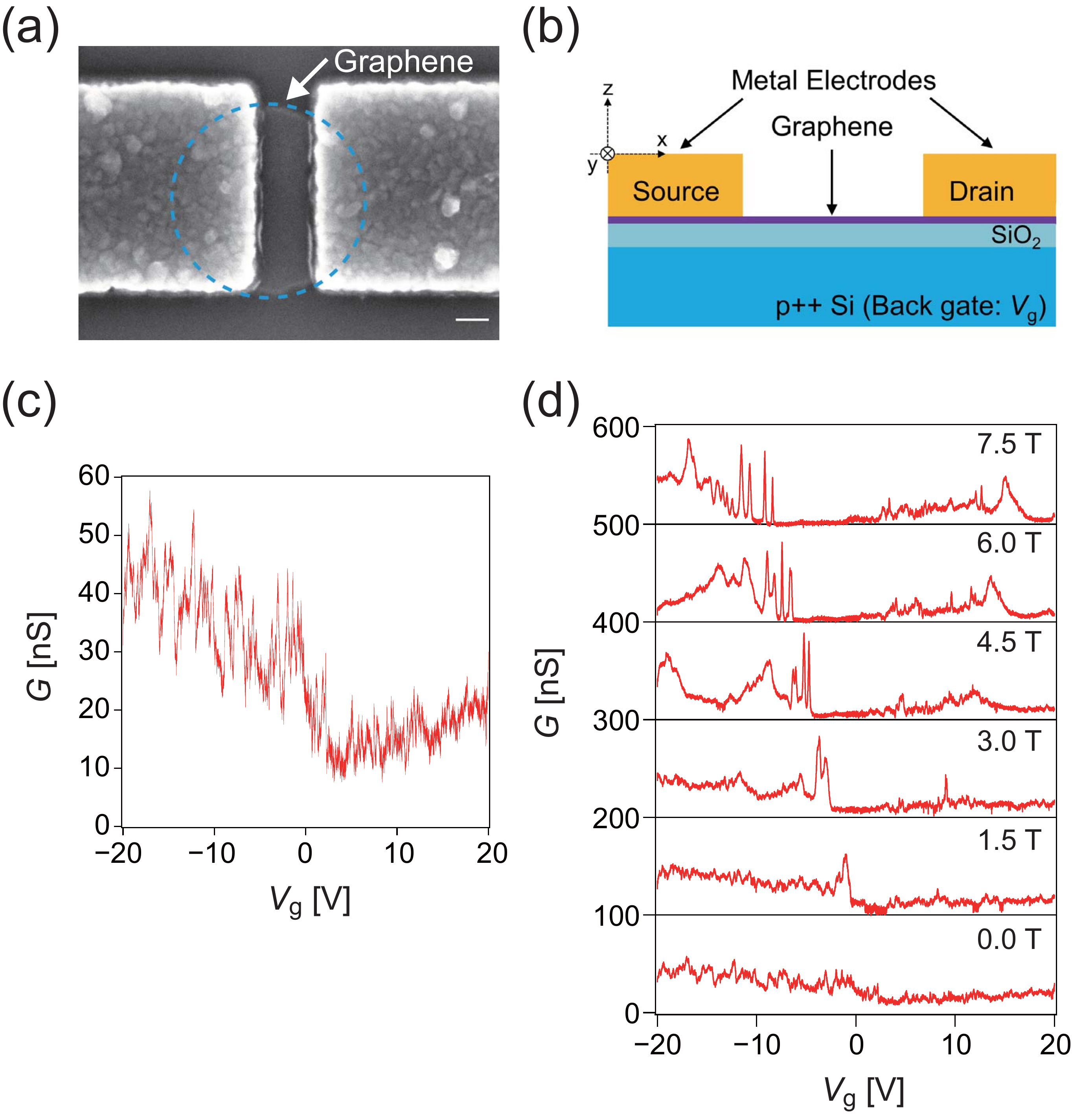}%
\caption{\label{fig1} (color). Configuration and transport characteristics of graphene device. 
(a) Scanning microscope image of the measured device with electrode assignment. The scale bar is 100 nm. (b) Side-view schematic of the device with an electrode. The axis directions are shown. The midpoint between the metal contacts is defined as $x = 0$, and the center of the width of the metal segment is defined as $y = 0$. (c) Conductance (measured at $V_{\text{sd}} = 100$ $\mu \text{V}$) as a function of $V_{\text{g}}$ at $T = 1.7$ K and $B = 0$ T. (d) Conductance as a function of $V_{\text{g}}$ under magnetic fields applied perpendicular to the graphene device at $T = 1.7$ K and $V_{\text{sd}} = 100$ $\mu \text{V}$. For clarity, each trace is shifted in 100 nS steps from the bottom to the top.
}
\end{figure}

Figure 1(a) shows a scanning electron microscopic image of a single-layer graphene device. The disk diameter is 550 nm, and the distance between contacts is $L = 200$ nm. A side-view schematic of the device is shown in Fig. 1(b). A highly-\textit{p}-doped silicon substrate was used as the back gate, with a 90 nm oxidized silicon layer serving as the gate dielectric. To measure their electrical properties, the graphene devices were mounted in an ${}^{3}\text{He}$ cryostat equipped with a superconducting magnet. A magnetic field ($B$) was applied perpendicular to the sample chip. All electrical-transport measurements were carried out at base temperatures of 0.23--1.7 K. The two-terminal conductance $G$ through the graphene device was measured as a function of the gate voltage $V_{\text{g}}$ by applying a fixed DC bias voltage $V_{\text{sd}}$. In addition, we measured the source-drain current $I$ as a function of both $V_{\text{sd}}$ and $V_{\text{g}}$ and obtained the differential conductance $dI/dV_{\text{sd}}$ by numerically differentiating the $I$-$V_{\text{sd}}$ curve. Figure 1(c) shows the conductance through the graphene mesoscopic device as a function of $V_{\text{g}}$ at $V_{\text{sd}} = 100$ $\mu \text{V}$, $B = 0$ T and $T = 1.7$ K. The conductance curve is asymmetric and V-shaped with fluctuations, though we consider that the very low conductance is due to poor contact between the metal (titanium) and the graphene \cite{XiaNNanotech11}. 
This point has been discussed also in ref.\cite{MoriyamaMEX13}.
It has a broad minimum around the Dirac point at $V_{\text{g}} = V_{\text{Dirac}} \sim + 4$ V. This asymmetry and shift in the gate voltage of the Dirac-point energy have been reported in previous studies \cite{XiaNNanotech11,HuardPRB08,XiaNanoLett09} and is referred to as the `hole-doping case'. It has been interpreted as resulting from pinning of the graphene hole-carrier density under the metal and the formation of the surface-potential profile by charge transfer across the metal/graphene junction. Therefore, as $V_{\text{g}}$ is varied, the band profile changes from $pp'p$ to $pn'p$, where $p$ ($n$) and $p'$ ($n'$) refer to positive (negative) charge densities in the region under the metal segment or between the source and drain electrodes, respectively, as a function of the gate voltage. As a result, when $V_{\text{g}} < V_{\text{Dirac}}$ ($V_{\text{g}} > V_{\text{Dirac}}$), a $pp'p$ ($pn'p$) junction is formed \cite{XiaNNanotech11,HuardPRB08,XiaNanoLett09}. 

Figure 1(d) shows how the conductance evolves in a magnetic field as a function of $V_{\text{g}}$, with $V_{\text{sd}} = 100$ $\mu \text{V}$ and $T = 1.7$ K. Near the Dirac point, conductance is strongly suppressed by the magnetic field. Furthermore, several resonance peaks emerge in the hole- and electron-carrier regions. In particular, clear resonance peaks are observed in the hole-carrier region ($V_{\text{g}} \lesssim V_{\text{Dirac}}$). These peaks correspond to the Coulomb-blockade effect, which is discussed below. The fluctuations in conductance as a function of $V_{\text{g}}$ are stable over time and reproducible. An enlarged view of the conductance as a function of $V_{\text{g}}$ and $B$ is shown in Fig. 3(a) for $V_{\text{sd}} = 500$ $\mu \text{V}$ and $T = 0.23$ K. 

\begin{figure}
\includegraphics[width=8cm, bb=0 0 657 918]{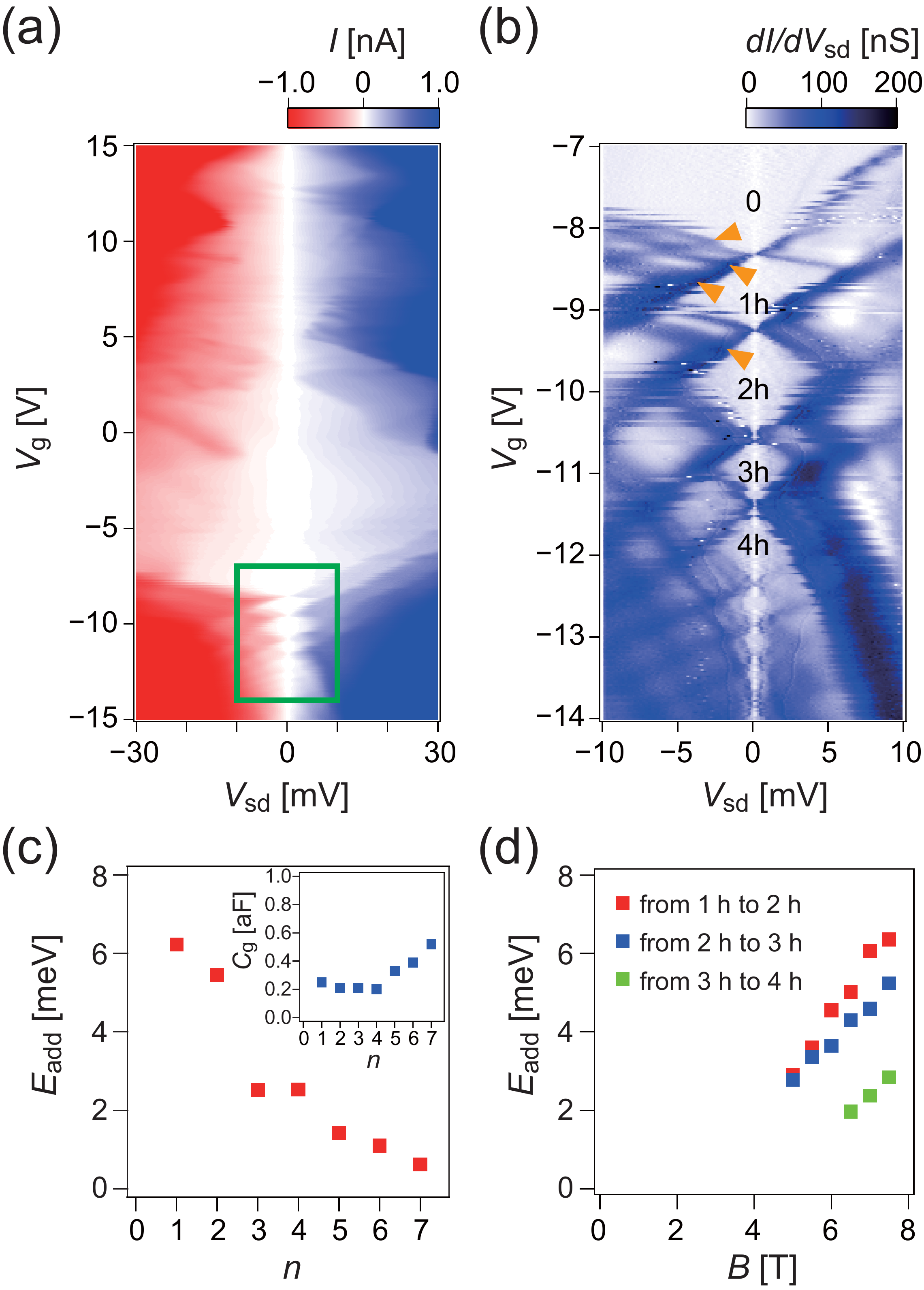}%
\caption{\label{fig2} (color). Coulomb blockade and quantum confinement under magnetic fields. 
(a) Large-scale current mapping as a function of $V_{\text{sd}}$ and $V_{\text{g}}$ at $T = 0.23$ K and $B = 7.5$ T. (b) High-resolution measurement of the differential conductance $dI/dV_{\text{sd}}$ as a function of $V_{\text{sd}}$ and $V_{\text{g}}$ for the green-boxed region of (a). Orange arrows point toward lines running parallel to the diamond edges, which correspond to the excitation spectra. The number of holes is fixed in each diamond region, and the `0, 1h, 2h, $\cdots$' labels indicate the number of hole carriers in the first Landau level (also see the text). (c) Addition energy $E_{\text{add}}$ as a function of hole number at $B = 7.5$ T. Inset shows the gate capacitance $C_{\text{g}}$ as a function of hole number at $B = 7.5$ T. (d) Magnetic field dependence of the addition energy, which is deduced from the size of the diamonds at each fixed magnetic field.
}
\end{figure}

To begin our discussion, we focus on the hole-carrier region. Figure 2(a) shows a large-scale plot of the source-drain current ($I$) as a function of both $V_{\text{sd}}$ and $V_{\text{g}}$ at fixed $B = 7.5$ T and $T = 0.23$ K. Diamond-shaped regions appear in the hole-carrier regions, which indicates that the current is suppressed due to Coulomb blockade. Figure 2(b) shows a high-resolution measurement of the differential conductance $dI/dV_\text{sd}$ as a function of both $V_{\text{sd}}$ and $V_{\text{g}}$ that corresponds to the green-boxed region in Fig. 2(a). Characteristic features called `Coulomb diamonds' manifest themselves, and the excitation spectra also appear, as shown by the orange arrows in Fig. 2(b). 
These data allow us to estimate the single-particle energy-level spacings ($\Delta E$) to be $\approx$ 2 meV. The lines for the excited state are broader than thermal broadening, which might be because of energy-dependent coupling of the excited state in the dot to the lead \cite{SchnezAPL09}. 

Under a perpendicular magnetic field, a Landau level (LL) generally forms in the 2DEG system. The LL energies ($E_{N}$) of the single-layer graphene sheet are given by $E_{N} = \text{sgn}(N) \upsilon_{\text{F}} \sqrt{2e\hbar|N|B}$, where $e$ is the fundamental unit of charge, $\hbar = h/(2\pi)$ where $h$ is Planck's constant, $\upsilon_{\text{F}}$ is the Fermi velocity, and the integer $N$ is the electron ($N > 0$) or a hole ($N < 0$) LL index \cite{CastroRMP09}. Note that the $N = 0$ LL ($E_{0}$) is independent of magnetic field in a single-layer graphene; i.e. $E_{0}$ is fixed at the Dirac-point energy with the magnetic field. However, the $N = 0$ LL spectrum does not appear in our device, which is attributed to the fluctuating spatial-disorder potential near the Dirac point and the $pn'p$ band profile, as discussed below. In Fig. 2b, we show the Coulomb diamond associated with the ridge of the first LL of hole carriers, which is quantized into discrete single-particle levels. The `0, 1h, 2h, $\cdots$' in each diamond indicates the number of carriers in the first LL of hole carriers. 

\begin{figure}
\includegraphics[width=8cm, bb=0 0 594 808]{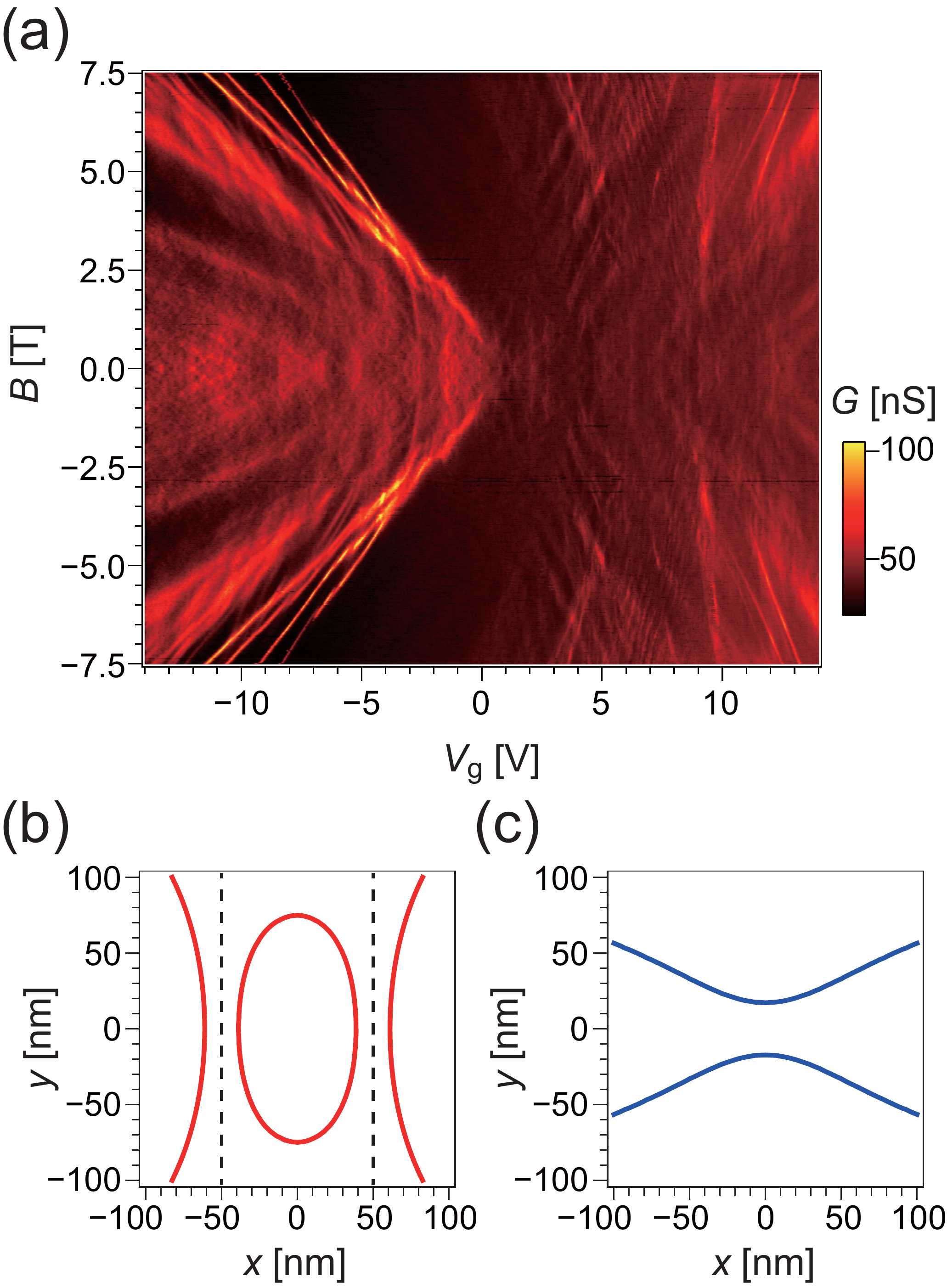}%
\caption{\label{fig3} (color). Field-induced quantum confinement-deconfinement in graphene.
(a) Conductance (for $V_{\text{sd}} = 500$ $\mu \text{V}$) mapping as a function of $V_{\text{g}}$ and $B$ at $T = 0.23$ K. 
(b) and (c) Typical classical trajectories of Dirac particle in a parabolic confinement potential. At a high magnetic field (b), closed orbits are approximately circular and coexist with open trajectories. As the magnetic field approaches the critical magnetic field ($B_{\text{c}}$), the orbits become increasingly elongated until finally merging with the separatrix (dashed black lines) at $B = B_{\text{c}}$. For a sufficiently weak magnetic field (c), all trajectories are open, which corresponds to a continuum spectrum. Based on the experimental data, we set the model parameters (see text) as follows: $U_{0} = 0.2$ eV ($V_{\text{g}} = - 10$ V), $B = 1$ (blue) and 2 T (red).
}
\end{figure}

We now focus on the `confinement-deconfinement transition' controlled by the magnetic field. As discussed in Ref. \cite{GuPRL11}, such a transition between open-to-closed trajectories of Dirac particles is ubiquitous in this type of device structure. Figs. 3(b) and (c) show the typical carrier trajectories between contacts in graphene with a confinement potential. For simplicity, we assumed that the electrostatic potential between contacts formed by the metal/graphene junction is a parabolic function \cite{GuPRL11,SilvestrovPRL07}: $U = -(4U_{0}/L^{2}) x^{2}$ where $U_{0} = \hbar \upsilon_{\text{F}}\sqrt{\pi C'_{\text{g}}|V_{\text{g}} - V_{\text{Dirac}}|/e}$ ($C'_{\text{g}}$ is the gate capacitance per area), which corresponds to the potential depth at a fixed gate voltage. Here we consider that electron motion in the $y$ direction is geometrically confined in our device. We set our model parameters on the basis of the measured area between the contacts of approximately 0.11 $\mu \text{m}^{2}$ ($= 0.2$ $\mu \text{m}$ $\times$ 0.55 $\mu \text{m}$) and $C'_{\text{g}} = 384$ aF $/ \mu \text{m}^{2}$. The solution for trajectory was obtained by numerically integrating the classical equation for a Dirac particle in a magnetic field. The qualitative features of our picture are not sensitive to the details of confining potential within the window of Figs. 3(b) and (c), so far as the potential is smooth under a magnetic field \cite{GuPRL11, Giavaras09, Maksym10}. As shown in Fig. 3(c), all the trajectories are open for a sufficiently weak magnetic field, which corresponds to a continuum spectrum. As the magnetic field approaches the critical magnetic field ($B_{\text{c}}$), the orbits asymptotically approach the critical separatrix lines which correspond to the dashed black lines in Fig. 3(b). The orbits finally merge with the separatrix at $B = B_{\text{c}}$. Above $B_{\text{c}}$, closed orbits emerge that correspond to quasi-bound states and coexist with open trajectories, as shown in Fig. 3(b). At high magnetic fields, the closed orbits are approximately circular. The quasi-bound states should be consistent with the Bohr-Sommerfeld condition \cite{GuPRL11, SilvestrovPRL07} and lead to a discrete energy spectrum. Furthermore, tunneling between open and closed trajectories leads to a finite resonance lifetime and QD formation \cite{SilvestrovPRL07,Giavaras09, Maksym10}. Resonance shapes are governed by charging effects via Coulomb blockade. Therefore, we consider that a genuine Dirac particle has been confined. In addition, the appearance of four clear Coulomb peaks, e.g. for the 7.5 T [see the hole-carrier regions in Fig. 1(d)], suggests two-valley degenerate quantum states with two-fold spin degeneracy \cite{WangNNanotech11, LiangPRL02, MoriyamaPRL05}.

Having understood the qualitative behavior of field-induced quantum dots in the graphene device, we quantitatively analyze the Coulomb-blockade effect to substantiate our picture and better understand the finer details. The main panel of Fig. 2(c) shows the addition energy ($E_{\text{add}}$) and gate capacitance of a single-electron charge ($C_{\text{g}}$) (inset) as a function of the number of holes. The constant-interaction model of a QD gives $E_{\text{add}} = E_{\text{C}} + \Delta E$, where $E_{\text{C}}= e^{2} / C_{\Sigma}$ is the charging energy and $C_{\Sigma}= C_{\text{s}} + C_{\text{d}} + C_{\text{g}}$, where $C_{\text{s}}$ and $C_{\text{d}}$ are the source and drain capacitances, respectively \cite{Kouwenhoven97}. Based on the experimental data shown in Fig. 2(b), we deduced $E_{\text{add}}$ from the diamond size and $C_{\text{g}} [= e / \Delta V_{\text{g}} (1 + \Delta E / E_{\text{C}})]$ from the gate-voltage spacing $\Delta V_{\text{g}}$ between diamonds at $V_{\text{sd}} = 0$. By setting the energy-level spacings $\Delta E \approx 0$, the classical capacitive-voltage relation for a single-electron charge becomes $C_{\text{g}} = e / \Delta V_{\text{g}}$. From the excitation spectra of Fig. 2(b), $\Delta E$ is estimated for the 1h and 2h Coulomb diamonds. Here $E_{\text{add}}$ and $C_{\text{g}}$ for 1h and 2h account for $\Delta E$, and the other $E_{\text{add}}$ and $C_{\text{g}}$ are deduced with $\Delta E \approx 0$. Although the addition energy $E_{\text{add}}$ decreases rapidly as holes are added, $C_{\text{g}}$ is nearly constant or slightly increases. We find that $C_{\text{g}}$ ($< 1$ aF) decreases by two or three orders of magnitude than the total capacitance (e.g. when $E_{\text{C}} = e^{2} / C_{\Sigma} = 4$ meV, the corresponding $C_{\Sigma} = 40$ aF), which implies that the capacitance between the lead and the QD is the main contributor to total capacitance. 
By using the $C_{\text{g}}$ data and the simple plate-capacitor model, the QD diameter is estimated to be 25--40 nm. These results are consistent with the above picture based on classical closed trajectory [Fig. 3(b)]. Figure 2(d) shows the magnetic field dependence of the addition energy, which is deduced from the size of diamond at each magnetic field value. The slope of each curve is $\sim 1$ meV/T; the slope cannot be attributed to the Zeeman effect, which would result in a slope of $g \mu_{\text{B}} = 0.12$ meV/T, assuming a $g$ factor of 2. The addition energy is mainly dominated by the charging energy, which in turn is dominated by the capacitance between the lead and the QD. Therefore, these results indicate that coupling between the lead and the QD is sensitive to the magnetic field, which is consistent with our scenario discussed above.
 
Finally, we consider the electron-carrier regions ($V_{\text{g}} \gtrsim V_{\text{Dirac}}$), in which the conductance peaks are irregular functions of the magnetic field [Fig. 1(d)]. In particular, peak crossings occur in Fig. 3(a), which implies transport through a multi-QD system. We consider that a multi-QD system forms as a result of the inhomogeneous $pn'p$ band profile along the contact metals (in the $y$ direction) and the spatial disorder potential fluctuations that arise because of the electron-hole puddle near the Dirac point \cite{MoriyamaMEX13, MartinNPhys08, JungNPhys11}. The formation of such a multi-QD system can also be deduced from the transport characteristics at a fixed magnetic field, as shown in Fig. 2(a). We observe a large transport gap ($\approx$ 20 meV) between the Dirac point and the 1h states in the ridge of the first LL, but the diamond structure collapses near the Dirac-point side and an irregular saw-tooth pattern is observed in the electron-carrier region.

In conclusion, we propose a graphene device structure that is not disturbed by the structural fluctuations of constrictions and that is directly connected to the mesoscopic 2DEG system. We demonstrated the Coulomb-blockade effect and quantum confinement, which are induced by both a uniform magnetic field perpendicular to the graphene sheet and an electrostatic surface-potential formed by the metal/graphene junction. Our experimental results indicate that a Dirac-type particle is confined and that quantum confinement-deconfinement transition occurs in the mesoscopic graphene system. 

We thank A. Kanda of Tsukuba University and K. Ishibashi of RIKEN for useful discussions. This study was supported by a Grant-in-Aid for Young Scientists (A), a Grant-in-Aid for Young Scientists (B), a Grant-in-Aid for Exploratory Research, the Nanotechnology Network Program, and the World Premier International Research Center Initiative on Materials Nanoarchitectonics from the Japan Ministry of Education, Culture, Sports, Science and Technology.

\end{document}